\PassOptionsToPackage{unicode}{hyperref}
\PassOptionsToPackage{hyphens}{url}
\PassOptionsToPackage{dvipsnames,svgnames,x11names}{xcolor}
\documentclass[
  12pt,
  letterpaper,
  DIV=11,
  numbers=noendperiod]{scrartcl}
\usepackage{xcolor}
\usepackage{amsmath,amssymb}
\usepackage{iftex}
\ifPDFTeX
  \usepackage[T1]{fontenc}
  \usepackage[utf8]{inputenc}
  \usepackage{textcomp} % provide euro and other symbols
\else % if luatex or xetex
  \usepackage{unicode-math} % this also loads fontspec
  \defaultfontfeatures{Scale=MatchLowercase}
  \defaultfontfeatures[\rmfamily]{Ligatures=TeX,Scale=1}
\fi
\usepackage{lmodern}
\ifPDFTeX\else
\fi
\IfFileExists{upquote.sty}{\usepackage{upquote}}{}
\IfFileExists{microtype.sty}{% use microtype if available
  \usepackage[]{microtype}
  \UseMicrotypeSet[protrusion]{basicmath} % disable protrusion for tt fonts
}{}
\makeatletter
\@ifundefined{KOMAClassName}{% if non-KOMA class
  \IfFileExists{parskip.sty}{%
    \usepackage{parskip}
  }{% else
    \setlength{\parindent}{0pt}
    \setlength{\parskip}{6pt plus 2pt minus 1pt}}
}{% if KOMA class
  \KOMAoptions{parskip=half}}
\makeatother
\makeatletter
\ifx\paragraph\undefined\else
  \let\oldparagraph\paragraph
  \renewcommand{\paragraph}{
    \@ifstar
      \xxxParagraphStar
      \xxxParagraphNoStar
  }
  \newcommand{\xxxParagraphStar}[1]{\oldparagraph*{#1}\mbox{}}
  \newcommand{\xxxParagraphNoStar}[1]{\oldparagraph{#1}\mbox{}}
\fi
\ifx\subparagraph\undefined\else
  \let\oldsubparagraph\subparagraph
  \renewcommand{\subparagraph}{
    \@ifstar
      \xxxSubParagraphStar
      \xxxSubParagraphNoStar
  }
  \newcommand{\xxxSubParagraphStar}[1]{\oldsubparagraph*{#1}\mbox{}}
  \newcommand{\xxxSubParagraphNoStar}[1]{\oldsubparagraph{#1}\mbox{}}
\fi
\makeatother

\usepackage{longtable,booktabs,array}
\usepackage{calc} % for calculating minipage widths
\usepackage{etoolbox}
\makeatletter
\patchcmd\longtable{\par}{\if@noskipsec\mbox{}\fi\par}{}{}
\makeatother
\IfFileExists{footnotehyper.sty}{\usepackage{footnotehyper}}{\usepackage{footnote}}
\makesavenoteenv{longtable}
\usepackage{graphicx}
\makeatletter
\newsavebox\pandoc@box
\newcommand*\pandocbounded[1]{% scales image to fit in text height/width
  \sbox\pandoc@box{#1}%
  \Gscale@div\@tempa{\textheight}{\dimexpr\ht\pandoc@box+\dp\pandoc@box\relax}%
  \Gscale@div\@tempb{\linewidth}{\wd\pandoc@box}%
  \ifdim\@tempb\p@<\@tempa\p@\let\@tempa\@tempb\fi% select the smaller of both
  \ifdim\@tempa\p@<\p@\scalebox{\@tempa}{\usebox\pandoc@box}%
  \else\usebox{\pandoc@box}%
  \fi%
}
\def\fps@figure{htbp}
\makeatother

\NewDocumentCommand\citeproctext{}{}
\NewDocumentCommand\citeproc{mm}{%
  \begingroup\def\citeproctext{#2}\cite{#1}\endgroup}
\makeatletter
 \let\@cite@ofmt\@firstofone
 \def\@biblabel#1{}
 \def\@cite#1#2{{#1\if@tempswa , #2\fi}}
\makeatother
\newlength{\cslhangindent}
\newlength{\csllabelwidth}
\newenvironment{CSLReferences}[2] % #1 hanging-indent, #2 entry-spacing
 {\begin{list}{}{%
  \setlength{\itemindent}{0pt}
  \setlength{\leftmargin}{0pt}
  \setlength{\parsep}{0pt}
  \ifodd #1
   \setlength{\leftmargin}{\cslhangindent}
   \setlength{\itemindent}{-1\cslhangindent}
  \fi
  \setlength{\itemsep}{#2\baselineskip}}}
 {\end{list}}
\usepackage{calc}

\usepackage{tcolorbox}
\usepackage{amssymb}
\usepackage{yfonts}
\usepackage{bm}

\newtcolorbox{greybox}{
  colback=white,
  colframe=blue,
  coltext=black,
  boxsep=5pt,
  arc=4pt}

\KOMAoption{captions}{tableheading}
\makeatletter
\@ifpackageloaded{caption}{}{\usepackage{caption}}
\AtBeginDocument{%
\ifdefined\contentsname
  \renewcommand*\contentsname{Table of contents}
\else
  \newcommand\contentsname{Table of contents}
\fi
\ifdefined\listfigurename
  \renewcommand*\listfigurename{List of Figures}
\else
  \newcommand\listfigurename{List of Figures}
\fi
\ifdefined\listtablename
  \renewcommand*\listtablename{List of Tables}
\else
  \newcommand\listtablename{List of Tables}
\fi
\ifdefined\figurename
  \renewcommand*\figurename{Figure}
\else
  \newcommand\figurename{Figure}
\fi
\ifdefined\tablename
  \renewcommand*\tablename{Table}
\else
  \newcommand\tablename{Table}
\fi
}
\@ifpackageloaded{float}{}{\usepackage{float}}
\floatstyle{ruled}
\@ifundefined{c@chapter}{\newfloat{codelisting}{h}{lop}}{\newfloat{codelisting}{h}{lop}[chapter]}
\floatname{codelisting}{Listing}

\makeatother
\makeatletter
\@ifpackageloaded{caption}{}{\usepackage{caption}}
\@ifpackageloaded{subcaption}{}{\usepackage{subcaption}}
\makeatother
\usepackage{bookmark}
\IfFileExists{xurl.sty}{\usepackage{xurl}}{} % add URL line breaks if available
\hypersetup{
  pdftitle={Metric/Nonmetric Elastic MDS},
  pdfauthor={Jan de Leeuw},
  colorlinks=true,
  linkcolor={blue},
  filecolor={Maroon},
  citecolor={Blue},
  urlcolor={Blue},
  pdfcreator={LaTeX via pandoc}}

\title{Metric/Nonmetric Elastic MDS}
\author{Jan de Leeuw}
\date{November 24, 2025}
\begin{document}
\maketitle
\begin{abstract}
We present R and C implementations for metric (ratio) and non-metric
(ordinal) versions of Elastic MDS, the multidimensional scaling
technique proposed by McGee (\citeproc{ref-mcgee_66}{1966}). The R and C
versions are compared for speed, with the C version anywhere from 15 to
100 times as fast as the R version.
\end{abstract}

\renewcommand*\contentsname{Table of contents}
{
\hypersetup{linkcolor=}
\setcounter{tocdepth}{3}
\tableofcontents
}

\textbf{Note:} This is a working manuscript which will be
expanded/updated frequently. All suggestions for improvement are
welcome. All Rmd, tex, html, pdf, R, and C files are in the public
domain. Attribution will be appreciated, but is not required. The files
can be found at \url{https://github.com/deleeuw/elastic}

\section{Introduction}\label{introduction}

Early in the history of computerized Multidimensional Scaling (MDS),
between the seminal contributions of Shepard/Kruskal and
Guttman/Lingoes/Roskam, there was the ``elastic method'' proposed by
Victor E. McGee in a series of papers (McGee
(\citeproc{ref-mcgee_65}{1965}), McGee (\citeproc{ref-mcgee_66}{1966}),
McGee (\citeproc{ref-mcgee_67}{1967}), McGee
(\citeproc{ref-mcgee_68}{1968})). The method has been largely forgotten,
but it is worth remembering, because it is different in some important
aspects from the more well-known methods.

The least squares loss function in most MDS methods (Borg and Groenen
(\citeproc{ref-borg_groenen_05}{2005})) can be written as
\begin{equation}
\sigma(X,\Delta):=\frac{\mathop{\sum\sum}_{1\leq i<j\leq n}w_{ij}(\delta_{ij}-d_{ij}(X))^2}{ \mathop{\sum\sum}_{1\leq i<j\leq n}w_{ij}\delta_{ij}^2},\label{eq-kruskal}
\end{equation} where \(X\) is the \emph{configuration} of \(n\)
\emph{points} in \(p\) \emph{dimensions}, \(d_{ij}(X)\) is the
\emph{Euclidean distance} between points \(i\) and \(j\) in
configuration \(X\), \(\delta_{ij}\) is the \emph{dissimilarity} between
points \(i\) and \(j\), and \(w_{ij}\) is a \emph{weight}. Weights are
non-negative.

In the \emph{metric} version of the MDS method the dissimilarities are
observed and fixed, and minimization is over configurations only. The
denominator in \eqref{eq-kruskal} is irrelevant for the minimization
problem, but it normalizes the problem in the sense that the minimum of
stress is between zero and one. In the \emph{nonmetric} version
minimization is over both configurations and dissimilarities, with the
constraint that the dissimilarities are monotonic with an observed set
of dissimilarities.

It is true that in the original Kruskal formulation the denominator is
the sum of the squared distances instead. But De Leeuw
(\citeproc{ref-deleeuw_U_75a}{1975}) shows that normalizing by using
either the sum of squared distances or the sum of squared
dissimilarities leads to the same solution (up to a scalar
proportionality factor). The smacof program (De Leeuw and Mair
(\citeproc{ref-deleeuw_mair_A_09c}{2009}), Mair, Groenen, and De Leeuw
(\citeproc{ref-mair_groenen_deleeuw_A_22}{2022})) minimizes the
numerator of \eqref{eq-kruskal} over configurations and dissimilarities,
with the additional constraint that the sum of squares of the
dissimilarities is equal to a constant. Again, this \emph{explicit
normalization} gives the same solution, up to proportionality, as the
original Kruskal formulation that uses \emph{implicit normalization} by
dividing by the sum of squared distances.

In McGee's elastic method the loss is constructed to satisfy two
requirements. The first one is

\begin{quote}
Psychological judgments which indicated relatively great separation of
stimuli should be allowed greater error than judgments indicating close
proximity (l.c. p.~182)
\end{quote}

And the second requirement is the basic MDS requirement that
dimensionality of \(X\) must as low as possible, while still providing a
good fit.

\begin{quote}
A criterion which suggested itself in response to the first requirement
was one based on the physical work done on an elastic spring to stretch
or compress it from an initial length \(\delta_{ij}\) to a final length
\(d_{ij}\). (l.c. p.~183)
\end{quote}

This leads to the loss function \begin{equation}
\sigma(X,\Delta):=\mathop{\sum\sum}_{1\leq i<j\leq n}w_{ij}\frac{(\delta_{ij}-d_{ij}(X))^2}{\delta_{ij}^2},\label{eq-mcgee}
\end{equation} which McGee calls \emph{work}. We shall just call it
\emph{stress} (and \emph{elastic stress} or \emph{McGee stress}), using
the more familiar MDS name for loss. We will also continue to use the
symbol \(\sigma\) for any least squares MDS loss function.

The elastic MDS problem is minimization of stress over both \(X\) and
\(\Delta\), where \(\Delta\) must be monotone with the given
dissimilarities. In \eqref{eq-mcgee} the weight \(w_{ij}\) is
interpreted as the modulus of elasticity of the spring \((i,j)\).

In McGee (\citeproc{ref-mcgee_67}{1967}) the alternative loss function
\begin{equation}
\sigma(X,\Delta):=\mathop{\sum\sum}_{1\leq i<j\leq n}w_{ij}\frac{(\delta_{ij}-d_{ij}(X))^2}{d_{ij}^2(X)},\label{eq-mcgee2}
\end{equation} is proposed. Minimizing \eqref{eq-mcgee2} seems more
complicated, and we will postpone studying algorithms to minimize it.
Thus we will work with \eqref{eq-mcgee} in this paper.

In McGee's papers the actual algorithm and its implementation are not
described in sufficient detail. Part of the problem is that he is
dealing exclusively with the nonmetric case, in which minimization over
both \(X\) and \(\Delta\) is necessary. In the metric case minimization
is over \(X\) only, and the algorithm is much simpler.

\section{Properties}\label{properties}

If we compare \eqref{eq-kruskal} and \eqref{eq-mcgee} we see one
important difference. The normalization in \eqref{eq-kruskal} is by the
sum of squared dissimilarities, while the normalization in
\eqref{eq-mcgee} is term-by-term, by the squared dissimilarities
themselves. An alternative way of writing the elastic loss function
makes this more clear. \begin{equation}
\sigma(X,\Delta):=\mathop{\sum\sum}_{1\leq i<j\leq n}w_{ij}\left(1-\frac{d_{ij}(X)}{\delta_{ij}}\right)^2\label{eq-mcgee3}
\end{equation} Thus we see that instead of minimizing the difference of
dissimilarities and distances from zero the elastic method minimizes the
deviations of their ratios from one.

If \(\delta_{ij}\) is multiplied by a positive constant, then so is the
optimum configuration \(X\) and the \(D(X)\). As a consequence, the
minimum of elastic stress does not change, i.e.~is homogeneous of degree
zero in the dissimilarities.

It is clear from \eqref{eq-mcgee3} that the loss function is undefined
if one or more of the dissimilarities are zero. If one of the distances
is zero, then the corresponding term in the loss function is equal to
the corresponding weight, irrespective of what the corresponding
dissimilarity is. Thus the minimum of elastic stress is always less than
or equal to the sum of the weights, its value at \(X=0\).

If \(\delta_{ij}\) and \(d_{ij}(X)\) are non-zero and close then a first
order Taylor series expansion gives \begin{equation}
\log d_{ij}(X)-\log\delta_{ij}\approx\frac{1}{\delta_{ij}}(d_{ij}(X)-\delta_{ij}),\label{eq-ddelta}
\end{equation} from which it follows that \begin{equation}
\sigma(X,\Delta)\approx\mathop{\sum\sum}_{1\leq i<j\leq n}w_{ij}(\log\delta_{ij}-\log d_{ij}(X))^2.\label{eq-approx}
\end{equation} If the fit is good the elastic stress will be
approximately equal to the logarithmic stress from Ramsay
(\citeproc{ref-ramsay_77}{1977}). Or, to put it differently, minimizing
elastic stress can serve as an approximation to minimizing logarithmic
stress.

\section{Algorithm}\label{algorithm}

In this paper we will give an iterative algorithm for minimizing loss
from \eqref{eq-mcgee} in the metric and non-metric (ordinal) case. A
majorization algorithm for the metric case is already available in the
smacofx package (Rusch et al.
(\citeproc{ref-rusch_deleeuw_chen_mair_25}{2025}), Rusch et al.
(\citeproc{ref-rusch_deleeuw_mair_hornik_25}{In Press})). Our new
non-metric algorithm is in the \emph{alternating least squares} family,
which means we alternate minimizing over \(X\) for fixed \(\Delta\) and
over \(\Delta\) for fixed \(X\).

We will start our iterations with the classical metric solution
(Torgerson (\citeproc{ref-torgerson_58}{1958})) for \(\Delta\). We
actually scale that solution by minimizing \begin{equation}
\sigma(\lambda):=\sum_{k=1}^m \frac{w_k}{\delta_k^2}(\delta_k-\lambda d_k(X))^2\label{eq-lstress}
\end{equation} over \(\lambda\). The minimum is attained at
\begin{equation}
\hat\lambda:=\frac{\sum_{k=1}^m \frac{w_k}{\delta_k}d_k(X)}{\sum_{k=1}^m \frac{w_k}{\delta_k^2}d_k^2(X)}.\label{eq-lbd}
\end{equation}

In iteration \(k\) we perform one majorization step to replace
\(X^{(k)}\) by \(X^{(k+1)}\) to improve loss for fixed \(\Delta^{(k)}\)
and one monotone regression step to replace \(\Delta^{(k)}\) by
\(\Delta^{(k+1)}\) for fixed \(X^{(k+1)}\).

To minimize \eqref{eq-mcgee} over \(X\) for fixed \(\Delta\) we rewrite
loss as \begin{equation}
\sigma(X,\Delta)=\mathop{\sum\sum}_{1\leq i<j\leq n}\frac{w_{ij}}{\delta_{ij}^2}(\delta_{ij}-d_{ij}(X))^2.\label{eq-mcgee4}
\end{equation} This can be minimized (or decreased) by using the
standard smacof majorization step with the weights
\(w_{ij}/\delta_{ij}^2\).

To minimize over \(\Delta\) for given \(X\) we define
\(\gamma_{ij}:=-\delta_{ij}^{-1}\) abd \(c_{ij}(X):=-d_{ij}^{-1}(X)\).
Rewrite loss as \begin{equation}
\sigma(X,\Gamma)=\mathop{\sum\sum}_{1\leq i<j\leq n}w_{ij}d_{ij}^2(X)(\gamma_{ij}-c_{ij}(X))^2,\label{eq-mcgee5}
\end{equation} which we must minimize over increasing \(\gamma_{ij}\).
This is just monotone regression with the \(c_{ij}(X)\) as the targets
and with weights \(w_{ij}d_{ij}^2(X)\). After we have found the optimal
\(\hat\gamma_{ij}\) we transform back to
\(\hat\delta_{ij}=-\hat\gamma_{ij}^{-1}\).

\section{Software}\label{software}

The github repository contains R and C versions of the smacofSSElastic
program. Both smacofSSElasticC() and smacofSSElasticR() have the same
default values of the parameters and the two programs are structured the
same way. We iterate a maximum of 1000 iterations until the change in
stress from one iteration to the next is less than \ensuremath{10^{-6}}.
A single majorization update is alternated with a single monotone
regression.

smacofSSElasticC() has a driver in R that sets up the MDS data structure
(De Leeuw (\citeproc{ref-deleeuw_E_25b}{2025})). Both program use a
number of R utilities for data manipulation and the initial
configuration. One important difference is that the R version uses
gpava() for monotone regression (De Leeuw, Hornik, and Mair
(\citeproc{ref-deleeuw_hornik_mair_A_09}{2009})), which is written in R,
while the R version uses the C routine monotone() from Busing
(\citeproc{ref-busing_22}{2022}).

\section{Examples}\label{examples}

We analyze two classical MDS examples: the color data of Ekman
(\citeproc{ref-ekman_54}{1954}) and the Morse code data of Rothkopf
(\citeproc{ref-rothkopf_57}{1957}). We are only interested in this paper
in performance of the programs, not in the substantive interpretation of
the results or in the comparison of elastic scaling and regular smacof.

Numerical elastic scaling takes 586 iterations to arrive at stress
2.3268637. For ordinal we need 437 iterations for stress 0.056998. Since
the ordinal fit is very good the log-stress of \eqref{eq-approx} should
be close to the elastic stress. It is 0.0581521.

The results of a comparison of the R and C implementations with
microbenchmark (Mersmann (\citeproc{ref-mersmann_24}{2024})) yields the
times in microseconds in following table.

\begin{verbatim}
               min     lq   mean median     uq    max
R/Numerical  40.13  41.66  44.38  43.84  45.11 109.41
C/Numerical   1.44   1.53   1.60   1.60   1.64   1.81
R/Ordinal   163.52 168.14 172.10 169.50 172.28 242.27
C/Ordinal     4.06   4.39   4.57   4.46   4.71   7.37
\end{verbatim}

Using median times we see that that C version is about 40 times as fast
as the R version.

For the Morse code data numerical elastic scaling takes 212 iterations
to arrive at stress 64.3828862, for ordinal we need 152 iterations for
stress 29.7732277. Because the fit is much worse than in the Ekman case
we do not expect log-stress to be close to elastic stress. For the
ordinal case final log-stress is 42.8850382.

\begin{verbatim}
               min     lq   mean median     uq    max
R/Numerical  80.71  85.50  92.21  87.61  91.38 162.41
C/Numerical   6.38   6.62   6.77   6.71   6.80  10.96
R/Ordinal   711.81 744.80 767.76 764.32 784.45 862.41
C/Ordinal     7.25   7.60   7.82   7.74   7.83  11.35
\end{verbatim}

In the numerical case the C version is about 15 times as fast, in the
ordinal case about 100 times. The huge difference in the ordinal case is
likely to be due in large part to the different monotone regression
routiones used by the R and C programs.

Of course these microbenchmark results depend on the default parameters
of the programs (same in R and C), on the speed of my computer (Mac Mini
with Apple M4 Pro, 64 GB of Ram, Tahoe 26.2), and on my programming
habits (same in R and C).

\section*{References}\label{references}
\addcontentsline{toc}{section}{References}

\phantomsection\label{refs}

\end{document}